\documentclass[%
 reprint,
 amsmath,amssymb,
 aps,
]{revtex4-1}

\usepackage{graphicx}
\usepackage{dcolumn}
\usepackage{bm}

\begin{document}

\preprint{APS/123-QED}

\title{Lattice animals in diffusion limited binary colloidal system}
\thanks{A footnote to the article title}%

\author{Zakiya Shireen}
\author{Sujin B Babu}%
 \email{sujin@physics.iitd.ac.in}
\affiliation{Department of Physics, Indian Institute of Technology Delhi\\  Hauz Khas, New Delhi 110016,India.}%




\date{\today}

\begin{abstract}
In soft matter system controlling the structure of the amorphous materials have been a key challenge. In this work we have modeled irreversible diffusion limited cluster aggregation of binary colloids, which serves as a model for chemical gels. Irreversible aggregation of binary colloidal particles lead to the formation of percolating cluster of  one species or both species also called bigels. Before the formation of the percolating cluster the system form self similar structure defined by a fractal dimension. For a one component system when the volume fraction is very small the clusters are far apart from each other and  the system has a fractal dimension of $1.8$. Contrary to this we will show that for the binary system we observe the presence of lattice animals which has a fractal dimension of $2$ irrespective of the volume fraction. When the clusters start inter penetrating we observe a fractal dimension of $2.5$ same as in the case of one component system. We were also able to predict the formation of bigels using a simple inequality relation. We have also shown that the growth of clusters follows the kinetic equations introduced by Smoluchowski for diffusion limited cluster aggregation. Further more we are also proposing a universal parameter for irreversible binary colloidal system, which follows the scaling laws proposed by percolation theory. 
\end{abstract}

\pacs{Valid PACS appear here}
\maketitle


\section{Introduction}

	Irreversible aggregation of colloidal particles like proteins\cite{mezzenga2013self,zaccarelli2007colloidal}, DNA grafted nano-materials \cite{hecht2016kinetically,blumlein2015bigels,varrato2012arrested} etc leads to the formation of percolating clusters or gels. These amorphous materials are very important from a fundamental point as well  as find a range of application in industry\cite{boles2016self,duguet2011design}. The scaling laws associated with structure and kinetics of amorphous materials are very well explained by the percolation theory\cite{stauffer1979scaling}. Two limiting cases for irreversible aggregation are diffusion limited cluster aggregation (DLCA) and reaction limited cluster aggregation (RLCA) model both of which have been extensively studied \cite{meakin1983formation,botet1984hierarchical,klein1989universality,orrite2005off,lu2013colloidal}. The aggregation number $m$ of the self similar clusters formed from irreversible aggregation of particle is related to the radius of gyration $R_g$ by $m \propto R_g^{d_f}$ where $d_f$ is the fractal dimension of the clusters\cite{weitz1985limits}. Also the number density of the cluster $N(m)$ scale with the aggregation number as $N(m) \propto m^{-\tau}$ \cite{stauffer1979scaling}. In DLCA if the cluster are far apart from each other we have the flocculation regime characterized by $d_f=1.8$ and $\tau=0$, while when the clusters start to interpenetrate we have the percolation regime where $d_f=2.5$ and $\tau=2.2$. In reversible aggregation it has already been shown that the system undergoes a transition from continuous to directed percolation depending on strength of attraction \cite{kohl2016directed}.  

	Recently there have been many experimental and theoretical work on the  aggregation of binary colloids using patchy particles\cite{de2012bicontinuous}, DNA grafted on to the colloidal particles \cite{hecht2016kinetically,blumlein2015bigels,varrato2012arrested} or using two different types of micelles\cite{klymenko2014multiresponsive}, at finite temperatures. In the present work we have studied a model of irreversible DLCA for binary colloidal particles both having the same diameter and differ only in the way particles interact. The particles of the same species form irreversible bonds on collision, while particle of different species have hard core repulsive interaction. Depending on the fraction of species in the system we observe the appearance of percolated cluster of one or both species also called bigels.

	Here we report on the simulation study of DLCA binary colloidal system with short range interaction, where we have observed the appearance of bigels or one component gel depending on the fraction of each species for a particular volume fraction. We have proposed an inequality relation where by we were able to predict the appearance of bigel for a particular volume fraction. We have also shown that the aggregation kinetics of the system are very well described by the Smoluchowski rate equation. In the flocculation region instead of the fractal dimension of $d_f=1.8$ we observe the appearance of lattice animals which has a $d_f=2$, which we have also confirmed using the scaling between $N(m)$  and $m$. We have proposed a universal parameter for the species whose knetics is aressted and can not form percolating cluster. We have also shown that the universal parameter follows the scaling laws proposed by the percolation theory.

\section{Simulation}

	The simulation method used in the present work is called the Brownian cluster dynamics (BCD). BCD was introduced primarily to study the kinetics and dynamics of monomeric system. It has already been shown that this simulation techniques agree with the well known event driven molecular dynamics simulation\cite{babu2008influence}. In the present study we have modified BCD to accommodate binary spherical particles. We start our simulation with $N$ randomly distributed spheres of unit size  in a cubic box of length $L$. The volume fraction of the system is given by $\phi_{tot}=\frac{\pi}{6} N/L^3$. We randomly pick a fraction $c_A=N_A/N$ where $N_A$ is the number of $A$ particles and $c_B=1 -c_A$ is the fraction of $B$ particles present in the system. In the present study we will be working with different ratios $(c_A:c_B)$ of $A$ and $B$ particle, and we have always kept $c_B \geq c_A$.
	Inter species particles interact only via hard core repulsion, while intra species particles interact via a very short range square well potential with an interaction range of $0.1$ typical of a colloid. All the particles are displaced in a random direction using a step size $s$, and the time is incremented by $n s^2$ where $n$ is the number of simulation steps. It has already been demonstrated that if the step size $s$ is sufficiently small BCD is equivalent to Brownian dynamic simulation\cite{rottereau2005influence}. Thus $t=1$ is defined as the time taken by a monomer to travel its own diameter, where the diffusion coefficient of monomer is given by $D_0=1/6$. During random diffusion as soon as same kind of particles are within the range of the square well an irreversible bond is formed between them. All such connected spheres together are called a cluster, where a monomer is considered as a cluster of aggregation number $1$. After the diffusion of individual monomers within the cluster the center of mass of the cluster would have displaced in a random direction, on in other words we have followed Rouse dynamics\cite{rouse1953theory}. During the movement step if it leads to overlap with other sphere or leads to breaking of the bond we reject those movement steps. In addition to the Rouse dynamics the center of mass of the cluster is displaced in the same direction of the center of mass calculated from the Rouse dynamics. The displacement is now inversely proportional to its radius of the cluster \cite{babu2008influence} and if it leads to overlap with other cluster we reject the movement step mimicking Zimm dynamics \cite{zimm1956dynamics}.  The box size was varied from $L=50-100$ and the results in the present study is not influenced by finite size effects.

\section{Results}
\subsection{Phase diagram}
\begin{figure}
	\includegraphics[width=0.5\textwidth]{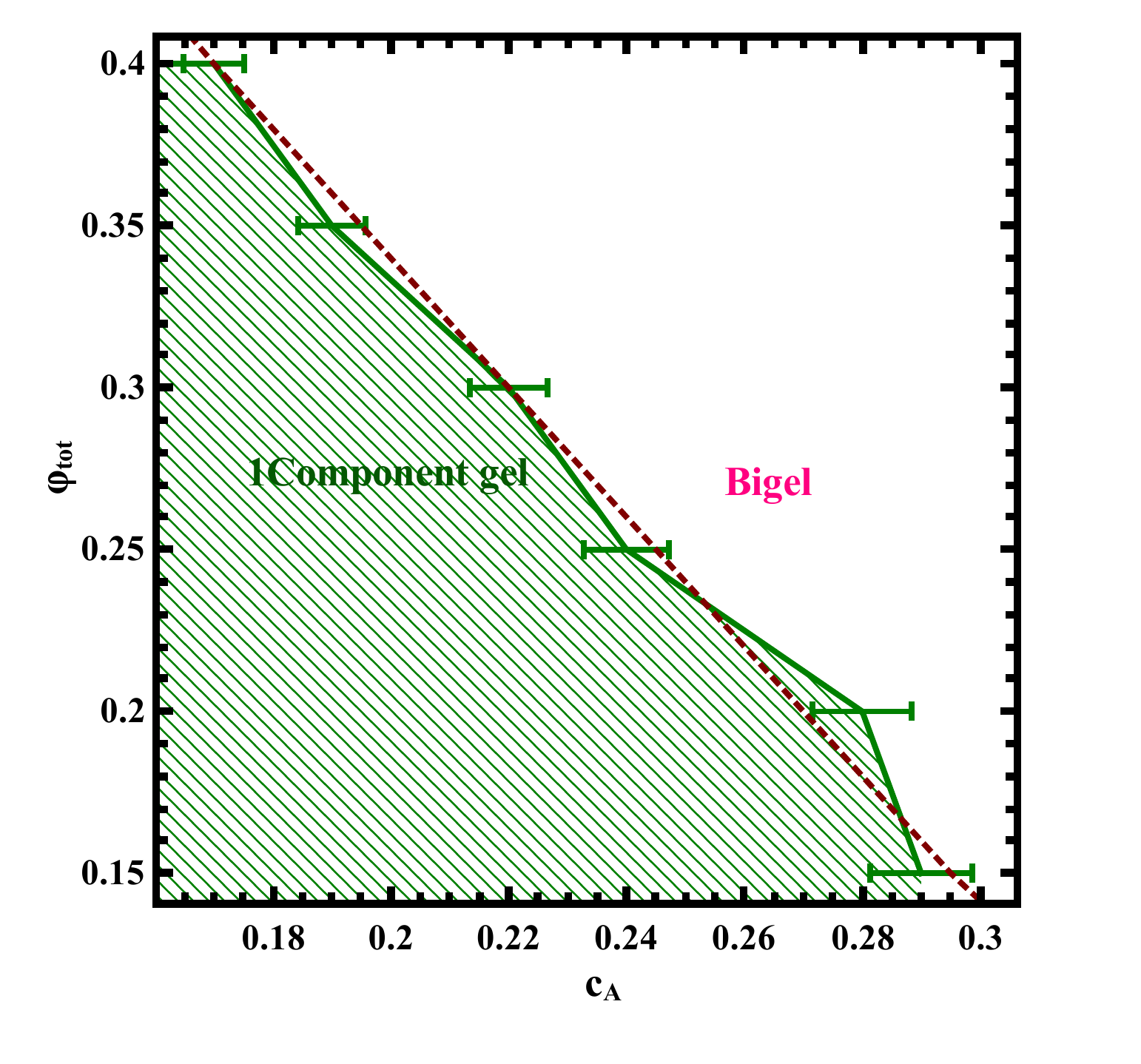}
	\caption{The volume fraction $\phi_{tot}$ is plotted against the fraction of $A$ particles $c_A$ . Below the curve shown in green we have only the $1$ component gel and above we have gels of both the particles or bigel. The error bar signifies two cases where we have observed the appearance of $1$ component gel and bigels more than $50\%$ in all the trials. The dashed line is given by $c_A=\frac{1}{2}(\phi_f-\phi_{tot})$.}
	\label{fig:1}
\end{figure}

	Irreversible aggregation of $1$ component colloidal system always leads to the formation of percolating clusters or gels \cite{meakin1983formation}, which in the present work we have defined as a cluster which extends between the opposite end of the simulation box. In the binary colloidal system as we change the fraction of $A$ particles $0<c_A<0.5$, we always observe a percolating cluster of the $B$ particles irrespective of the volume fraction, while the $A$ particles forms percolating cluster depending on the fraction of $c_A$. In Fig.\ref{fig:1} we have plotted $\phi_{tot}$ as a function of $c_A$, where we have identified regions as the one component gel when only $B$ particles percolate and $2$ component gel or bigel \cite{hecht2016kinetically,blumlein2015bigels,varrato2012arrested} when both the particles percolate. For identifying the $1$ component gel we have performed simulations for $10$ different configurations in a box size of $50$ at a particular combination of $c_A$ and $\phi_{tot}$.  If in more than $50\%$ of the trials only the $B$ particles formed a  percolating network and also the kinetics of $A$ particles is arrested we have defined it as a $1$ component gel. Likewise in $10$ trials if in more than $50\%$ of the trials the configuration resulted in a percolating cluster for both $A$ and $B$ particles we have defined it as a bigel. From Fig.\ref{fig:1}, we can observe that when $c_A<0.15$ we have a percolating cluster only for the $B$ particles irrespective of the $\phi_{tot}$. The aggregating $A$ clusters will not able to form a percolating cluster. The reason being the growth of the $A$ cluster is hindered by the presence of the $B$ percolating cluster. As we go to higher $c_A=0.18 \pm 0.01$ for $\phi_{tot}=0.4$, we have less particles in $B$ as well as more $A$ particles, which eventually results in the appearance of $A$ percolating cluster as there is less hindrance from the $B$ particles.  As we go to lower $\phi_{tot}$, the critical value of $c_A$ increases where we observe the formation of $A$ percolating network. 

	The transition from a $1$ component gel to bigel happens for different $\phi_{tot}$ at different $c_A$ values as is obvious from Fig.\ref{fig:1}. From the results of simulations for all the $\phi_{tot}$, we were able to deduce an upper bound when the bigel will appear in the system,
\begin{equation}
\frac{2 c_A}{\phi_f-\phi_{tot}} \geq 1
\label{eq1}
\end{equation}
	$\phi_f=0.74$ which also happens to be the maximum packing fraction for spheres. $\frac{ c_A}{\phi_f-\phi_{tot}}$ is the free space available in a randomly distributed spheres arrangement at a particular $\phi_{tot}$ for the $A$ particles. When the free volume available for the $A$ particles is approximately equal to half the volume of a sphere $\frac{2 c_A}{\phi_f-\phi_{tot}} \approx 1$, we observe that bigel start to appear in our system, as shown by the dashed line in Fig.\ref{fig:1}. When this fraction is less than one, the $A$ particles do not have enough space to diffuse and aggregate inside the pores of $B$ percolated cluster resulting in the formation of $1$ component gel. For very low $\phi_{tot}$ this relation may not be exactly followed and we may observe the appearance of bigel for $\frac{2 c_A}{\phi_f-\phi_{tot}}<1$. This is because the $A$ clusters can diffuse through the system as well as aggregate much more freely than for the higher $\phi_{tot}$, thereby equation \ref{eq1} only giving us an upper bound for the appearance of bigel.

\begin{figure}
	\includegraphics[width=0.5\textwidth]{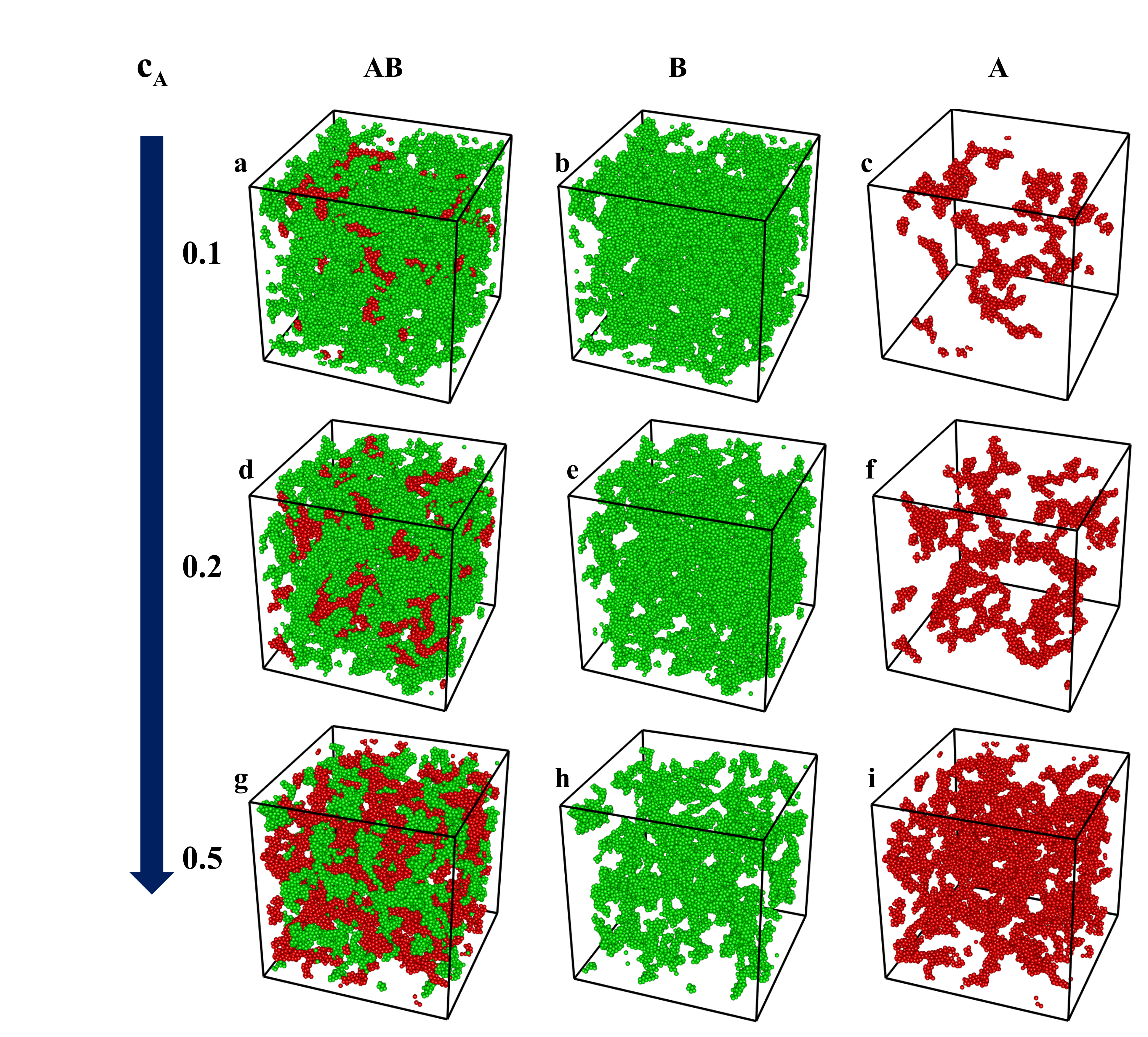}
	\caption{Snapshot of the binary system of particles where red represents $A$ particles and green represents $B$ particles for $\phi_{tot}=0.1$ in a box size $L=50$ for a range of increasing $c_A$ values as indicated by the arrow. {\bf(a)} Snapshot at $c_A=0.1$ of both $A$ and $B$ particles, {\bf(b)} for the $B$ particles after removing the $A$ particles where we have one single percolating cluster, {\bf(c)} for the $A$ particles after removing the $B$ particles where the small cluster are not able to grow due to the hindrance of $B$ percolating cluster. {\bf(d)} Snapshot at $c_A=0.2$ of the system of $A$ and $B$ particles {\bf(e)} $B$ particle alone where more free space is available and {\bf(f)} $A$ particle alone where we observe the presence of one big cluster and few smaller clusters distributed over the entire box. {\bf(g)} Snapshot at $c_A=0.5$ for both $A$ and $B$ particles, {\bf(h)} showing only the $B$ percolating cluster, which are more open compared to the earlier $2$ cases, {\bf(i)} showing only the $A$ particle which has aggregated into a single percolating cluster.}
	\label{fig:2}
\end{figure}
	In Fig.\ref{fig:2} we have shown snapshots of binary system for $3$ different $c_A$ as indicated on the left side of the figure for $\phi_{tot}=0.1$. The red colored spheres represent $A$ particle and green colored spheres represent $B$ particles. All the snapshot have been taken at a time when the aggregation kinetics of the system have stopped evolving. In Fig.\ref{fig:2}a we have shown both $A$ and $B$ particle as it appear in the system for $c_A=0.1$, evident from the fewer red particles seen in the system.  In Fig.\ref{fig:2}b we have kept only the $B$ particle , where we observe the presence of a single system spanning cluster, with thick tenuous branch. In Fig.\ref{fig:2}c we have shown only the $A$ particle for the same system which are forming  fractal like clusters distributed over the entire box. These small clusters are not able to aggregate further as they are stuck in the pores of $B$ percolating cluster. In Fig.\ref{fig:2}d we are showing both the particle for $c_A=0.2$, where we are closer to the critical point of $A$ particles, when it starts to percolate. In Fig.\ref{fig:2}e we have shown only the $B$ particle, where we observe that the gel of the $B$  looks much more open than in the case of  Fig.\ref{fig:2}b. As the number of $B$ particles is smaller than the previous system, the percolating cluster formed is at a lower volume fraction of $B$ particles. In the case of $A$ particles for the same $c_A$ we observe the presence of one very large cluster as well as smaller clusters distributed over the entire box. As we are close to the critical point we are able to observe clusters with  thin tenuous branches for the case of $A$ particles. The smaller clusters will never be able to form part of the percolating cluster as they are stuck inside the $B$ cluster. For the case of $c_A=0.5$ we are having the same number of $A$ and $B$ particle and both are forming independent system spanning clusters which are inter penetrating among each other see Fig.\ref{fig:2}g. In Fig.\ref{fig:2}h the $B$ percolating cluster are much more open than the previous $2$ cases as the percolating cluster is for a smaller number of $B$ particles compared to the earlier cases. The $A$ particles for $c_A=0.5$  are able to form percolating cluster with thicker strands and all the $A$ particles are now part of one single percolating cluster see Fig.\ref{fig:2}i. 

\begin{figure}
	\includegraphics[width=0.5\textwidth]{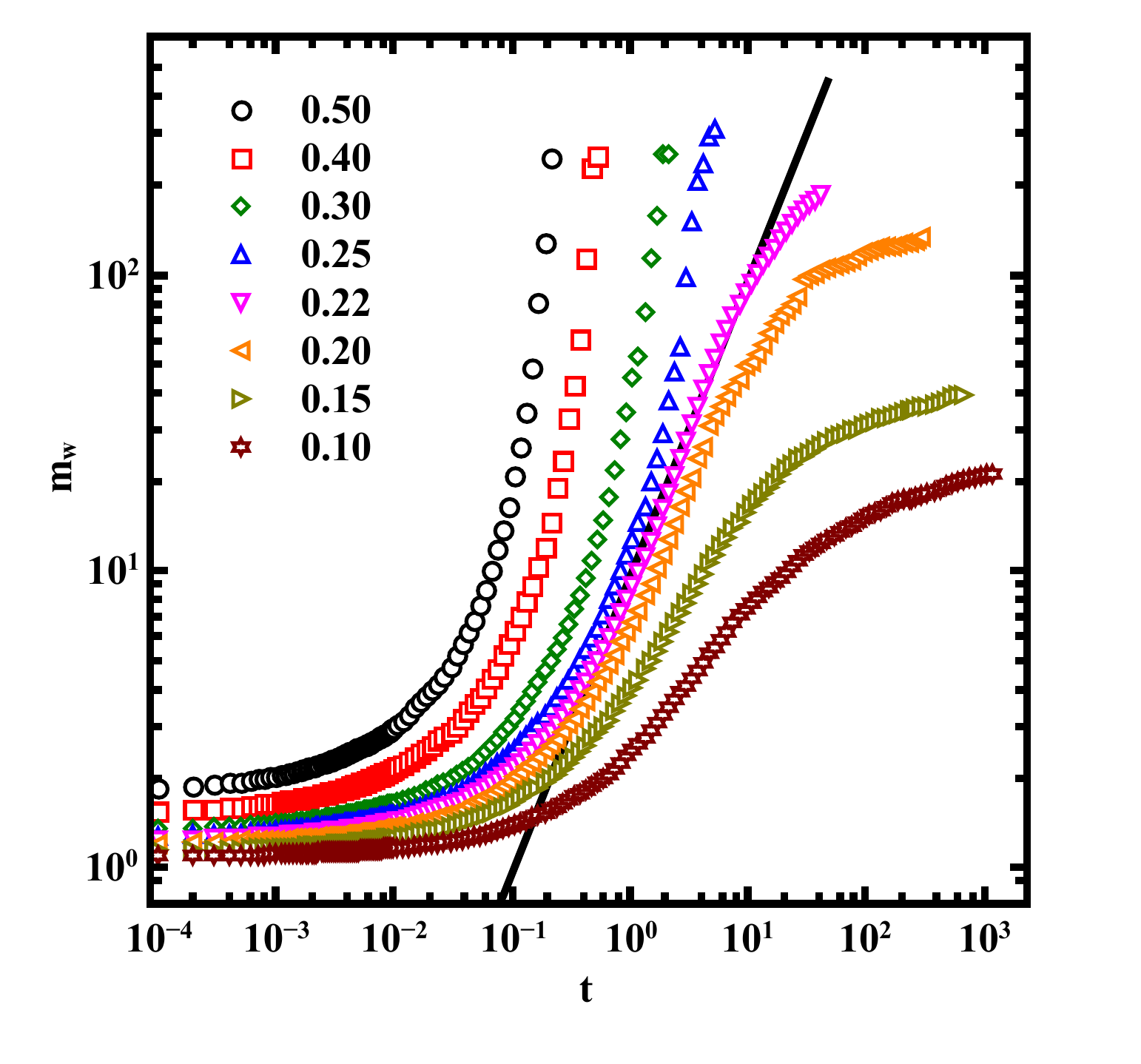}
	\caption{Mass average aggregation number ($m_w$) of only the $A$ particles is plotted as a function of time for range of $c_A$ at $\phi_{tot}=0.3$ as indicated in the figure. The straight line represents slope of unity in accordance with the Smoluchowski rate equation}
	\label{fig:3}
\end{figure}
\subsection{Kinetics of the $A$ particles}
	For studying the kinetics of aggregation we have calculated the mass average aggregation number $m_w= \sum_{m=1}^{\infty} m^2 N(m)/\sum_{m=1}^{\infty} mN(m)$ of the $A$ particles as a function of time for a range of $c_A$ at $\phi_{tot}=0.3$ see Fig.\ref{fig:3}. For $c_A>0.25$ we observe that $m_w$ diverges, indicating the formation of percolating cluster. For higher fraction $m_w$ diverges faster as there are more particles of $A$ species present in the system as well as $B$ cluster becomes much more open, thereby $A$ species will percolate faster. When $c_A<0.25$ we observe that $m_w$ grows for some time and then it stagnates, the reason being that the $A$ clusters are getting stuck inside the $B$ percolating cluster.  We know that in the flocculation limit the kinetics of DLCA type aggregation is very well explained by the Smoluchowski's rate equation where $m_w \propto t$ for the monomeric system \cite{von1917investigation,chandrasekhar1943stochastic}. The solid line in Fig.\ref{fig:3} has a slope of $1$, which is followed by all the fraction of $c_A$ in the flocculation limit. For $c_A>0.22$ we observe that the $m_w$ diverges from the slope of $1$, a signature of the formation of percolating cluster for the $A$ particles. For $c_A<0.22$ we observe that $m_w$ deviates to a smaller slope and at a later time approaches a stationary value indicating that $A$ particles are aggregating inside the cages of the $B$ clusters. As $c_A$ decreases the stationary value of $m_w$ also decreases as there is less space available for the $A$ particles to aggregate because of the increased fraction of $B$ particles. Although Smoluchowski approach is not valid in the percolation regime, as the clusters interpenetrate in this limit\cite{gimel1995transition}, we observe that for the present work it is valid up to the critical point. As $A$ clusters are caged inside the $B$ percolating cluster, the inter penetration  of $A$ clusters will be at a minimum which could be one possible reason why the $A$ clusters follow the Smoluchowski equation even close to the percolation limit. 
\begin{figure}
	\includegraphics[width=0.5\textwidth]{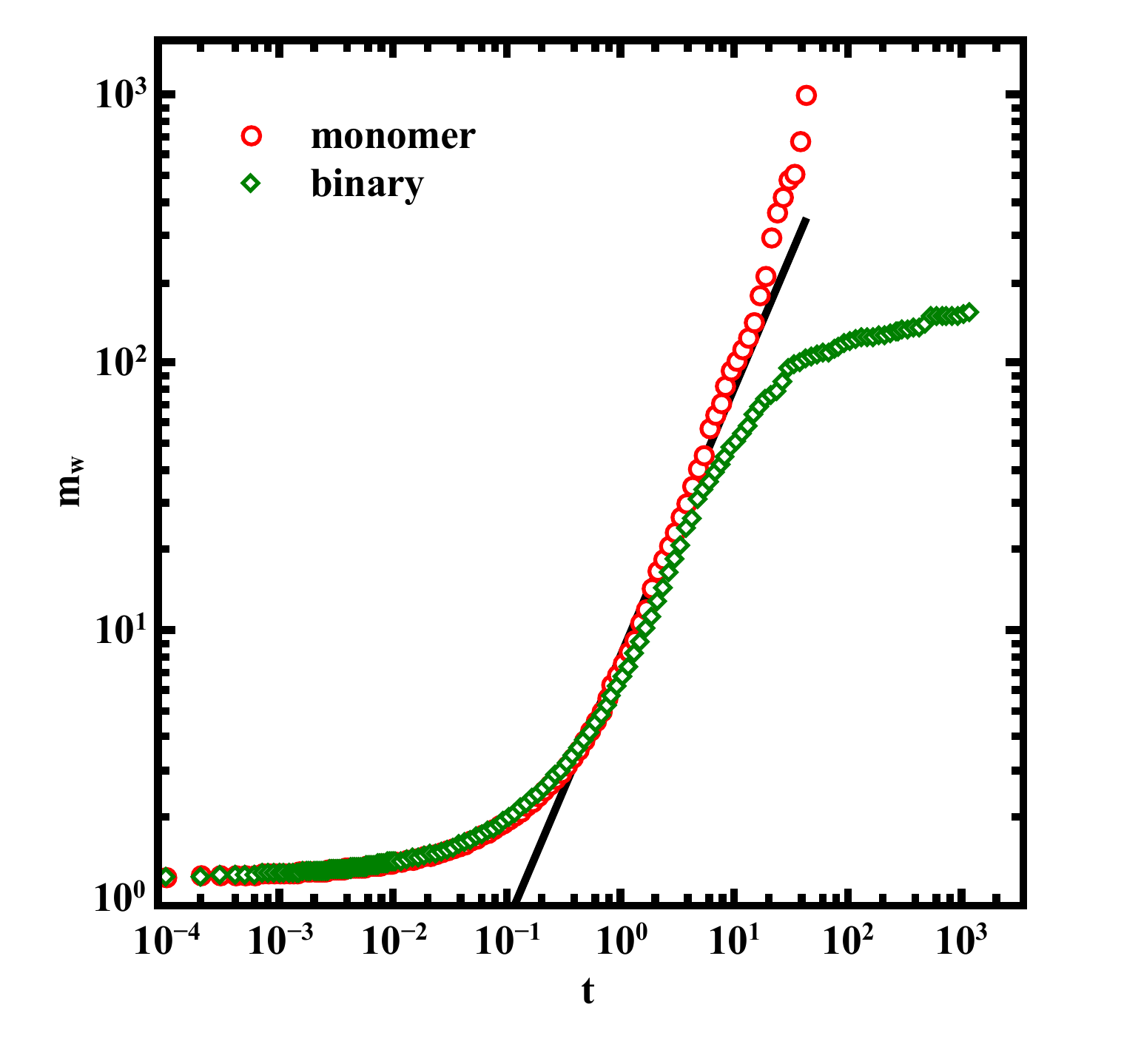}
	\caption{$m_w$ is plotted as a function of time for the $A$ particles  $c_A=0.2$ for the binary system (diamond) at $\phi_{tot}=0.3$ and monomeric system (circles) with volume fraction $0.06$. The solid line has a slope of $1$ according to the prediction of Smoluchowski rate equation \cite{chandrasekhar1943stochastic,von1917investigation}}
	\label{fig:4}
\end{figure}
\subsection{Kinetics of monomeric and binary system}
	It has already been shown that for monomeric system we will always form percolating cluster irrespective of the volume fraction\cite{gimel1995transition}. In Fig.\ref{fig:4} we have shown the evolution of $m_w$ for only $A$ particles when $\phi_{tot}=0.3$ at $c_A=0.2$ and for the monomeric system the volume fraction is $0.06$ the same as that for only $A$ particles in the system. For the initial time $t<1$ we observe the aggregation kinetics is following each other closely for the monomeric and binary case, as the monomers involved have displaced only approximately its own diameter. For $t>1$ we observe that the kinetics of aggregation for the binary system falls below a slope of $1$. While for the volume fraction $0.06$ the aggregation goes on with the slope $1$, which is expected from Smoluchowski approach. For the monomeric case $m_w$ deviate from a slope of unity for $t>10$ indicating the system have started to interpenetrate and later percolates at $t=59$. For the binary case $m_w$ approaches a stationary value see Fig.\ref{fig:4}. For all the volume fraction the monomeric case will form a percolating network while for the binary case the $A$ particles will form a percolating network only above a critical fraction for a particular $\phi_{tot}$. 
\begin{figure*}
	\includegraphics[width=1\textwidth]{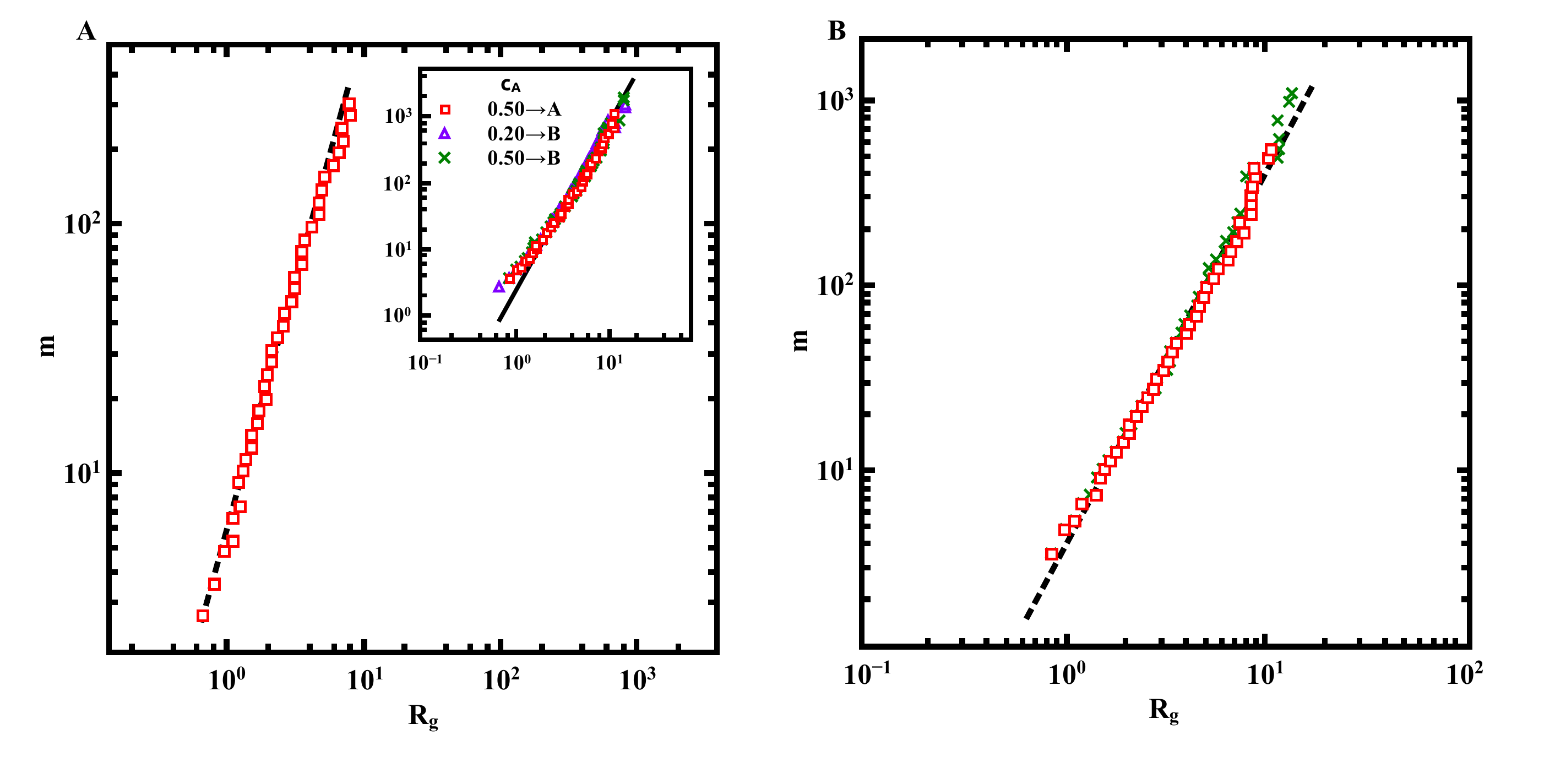}
	\caption{{\bf(a)} The aggregation number $m$ is plotted as function of $R_{g}$ is plotted for the $A$ particles for $c_A=0.2$ at $\phi_{tot}=0.2$. Here the $m_w$ attains a stationary value as $A$ clusters are stuck inside the percolated $B$ cluster. The dotted line has a slope of $2$, showing that the $A$ cluster is self similar and has a $d_f=2$. The inset shows the $m$ and $R_g$ for the A particles at $c_A=0.5$ (square) and for $B$ species at $c_A=0.5$ (cross) as well as $c_A=0.2$ (triangle) just before the percolated cluster appears. The solid line has a slope of $2.5$ showing that the system has already crossed over to percolation regime. {\bf(b)} Here $m$ and $R_g$ is plotted for $\phi_{tot}=0.15$ at $c_A=0.5$ for both $A$ (square) and $B$ (cross) during the early stage of aggregation. The dashed line has a slope of $2$ showing the clusters are still in the dilute regime with $d_f=2$.}
	\label{fig:5}
\end{figure*}
\subsection{Fractal dimension of binary system}
	For monomeric  DLCA type aggregation we know that there is a transition from flocculation regime to percolation regime, characterized by the change in fractal dimension from $1.8$ to $2.5$ \cite{gimel1995transition}. In Fig.\ref{fig:5}a we plot the aggregation number $m$ with the radius of gyration ($R_g$) of the $A$ clusters when $m_w$ attains a stationary value. The dotted line in Fig.\ref{fig:5}a has a slope of $2$, showing that the aggregation of $A$ cluster inside the cage of $B$ percolating cluster are self similar in nature with a $d_f=2$. In the inset of Fig.\ref{fig:5}a we have shown $m$ and $R_g$ of both $A$ and $B$ particles for $c_A=0.2$(triangle) and $c_A=0.5$(square for $A$ particles and cross for $B$ particles) before the system percolates. At $c_A=0.2$ $B$ particles are forming clusters and for $c_A=0.5$ both $A$ and $B$ species are forming self similar structure and have crossed over to the percolation regimes as shown in the inset of Fig.\ref{fig:5}a by the solid line which has a slope of $2.5$.  Usually $d_f=2$ is reported in reversible aggregation of colloids as well as during the formation of lattice animals \cite{aubert1986restructuring,schaefer1984fractal}. According to the percolation theory lattice animals appear far away from the percolation transition. Due to the presence of the $2$ species of particles there are many collisions between $A$ and $B$ particles, which do not lead to bond formation as they interact only through hard core repulsion. The clusters so formed are able to densify before a collision between same species happens leading to bond formation. This also leads to the conclusion that if $2$ species are undergoing irreversible aggregation, it leads to self similar clusters where all the configuration for a given aggregation number is equi-probable for both species of particles \cite{wessel1993cluster}in the flocculation regime. If the structure of the $A$ particles are influenced by the hindrance caused by the $B$ particles, we should be able to observe the same effect for the $B$ particles. This effect will be most evident when we have $c_A=0.5$, as we have same number of $A$ and $B$ particles. In Fig.\ref{fig:5}b we have plotted $m$ and $R_g$  for the case of $c_A=0.5 \; \phi_{tot}=0.15$ for the $A$ and $B$ in the initial stages of aggregation, where our system will still be in the flocculation regime. The solid line has a slope of $2$, which confirms the fact that for both $A$ and $B$ particles in flocculation regime have a $d_f=2$.  In the present case we were able to show that in the flocculation regime we have a $d_f=2$, which will cross over to $d_f=2.5$, as predicted by the percolation theory irrespective of type of particle. While in the case of $A$ particles when $m_w$ reaches a stationary value, it will always remain as lattice animals irrespective of $\phi_{tot}$	
\begin{figure*}
	\includegraphics[width=1\textwidth]{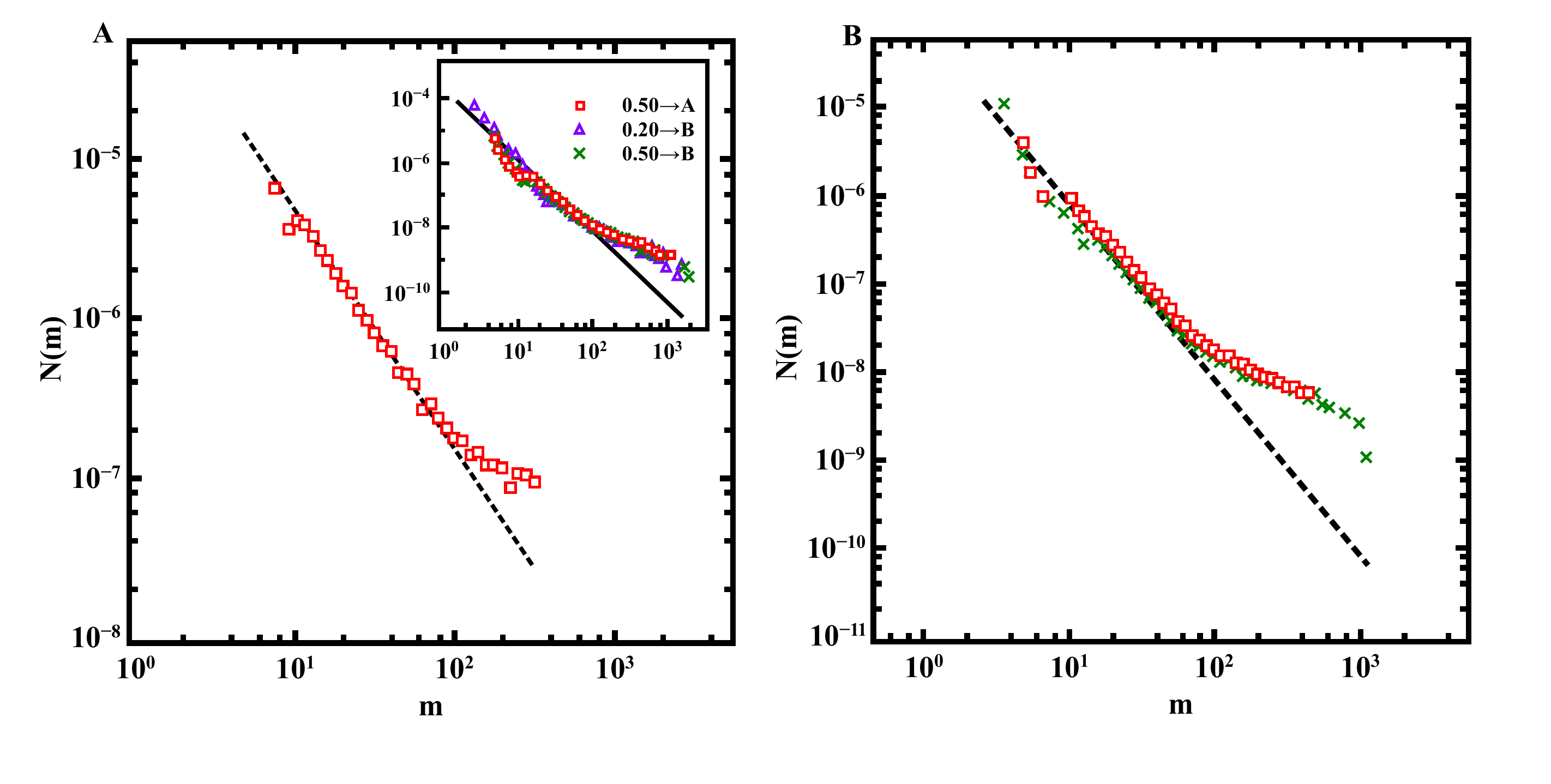}
	\caption{{\bf(a)} The number density of cluster $N(m)$ is plotted as a function of the aggregation number $m$ for $\phi_{tot}=0.2$ at $c_A=0.2$ for the $A$ clusters when $m_w$ attains a stationary value. The dashed line has a slope of $1.5$ as expected for lattice animals. The inset of figure we have plotted the $N(m)$ with aggregation number $m$ before the appearance of percolated cluster for the $A$ clusters at $c_A=0.5$ (square) and for the $B$ particles at $c_A=0.5$ (cross) and $c_A=0.2$ (triangle) for $\phi_{tot}=0.2$. The solid line in the inset has a slope of $2.2$ showing that the system is in percolation regime. {\bf(b)} $N(m)$ is plotted with $m$ for a system with $c_A=0.5$ at $\phi_{tot}=0.15$ during the initial stages of the aggregation process for the $A$ (square) and $B$ (cross) clusters. The dashed line has a slope of $1.5$ as predicted for lattice animals.}
	\label{fig:6}
\end{figure*}
\subsection{Scaling relation for binary system}
	 Conventional DLCA is characterized by $\tau=0$ for $d_f=1.8$ in the flocculation regime, for lattice animals we have $\tau=1.5$ and $d_f=2$, while in the percolation limit $\tau=2.2$ for $d_f=2.5$ as $N(m) \propto m^{-\tau}$ where the exponent $\tau$ are supposed to be exact \cite{stauffer1994introduction}. In Fig.\ref{fig:6}a, we have plotted $N(m)$ as a function $m$ for $\phi_{tot}=0.2$ at $c_A=0.2$ for the $A$ clusters, where $m_w$ attains a stationary value or in other words $A$ particles cannot percolate. The dotted line is given by a slope of $1.5$, as predicted for lattice animals by the percolation theory. While in the inset we have plotted $N(m)$ with $m$ for $B$ particles when $c_A=0.2$, also for both $A$ and $B$ particles at $c_A=0.5$ also see inset Fig.\ref{fig:5}a. The solid line which has a slope of $2.2$, which agrees with the fact that this system is already in the percolation regime. Also it shows that we have a cross over for the binary system from lattice animals to percolation regime. In Fig.\ref{fig:6}b we have plotted $N(m)$ with $m$ for $\phi_{tot}=0.15$ at $c_A=0.5$ both for $A$ and $B$ particles during the initial stages of aggregation process. The dotted line has a slope of $1.5$ confirming the fact that both $A$ and $B$ have a fractal dimension of $2$, contrary to the fact that in the flocculation regime for DLCA we have $d_f=1.8$. This we believe is a direct evidence that we are observing lattice animals in our system in the case when clusters are in the flocculation regime.
 
\begin{figure*}
 	\includegraphics[width=1\textwidth]{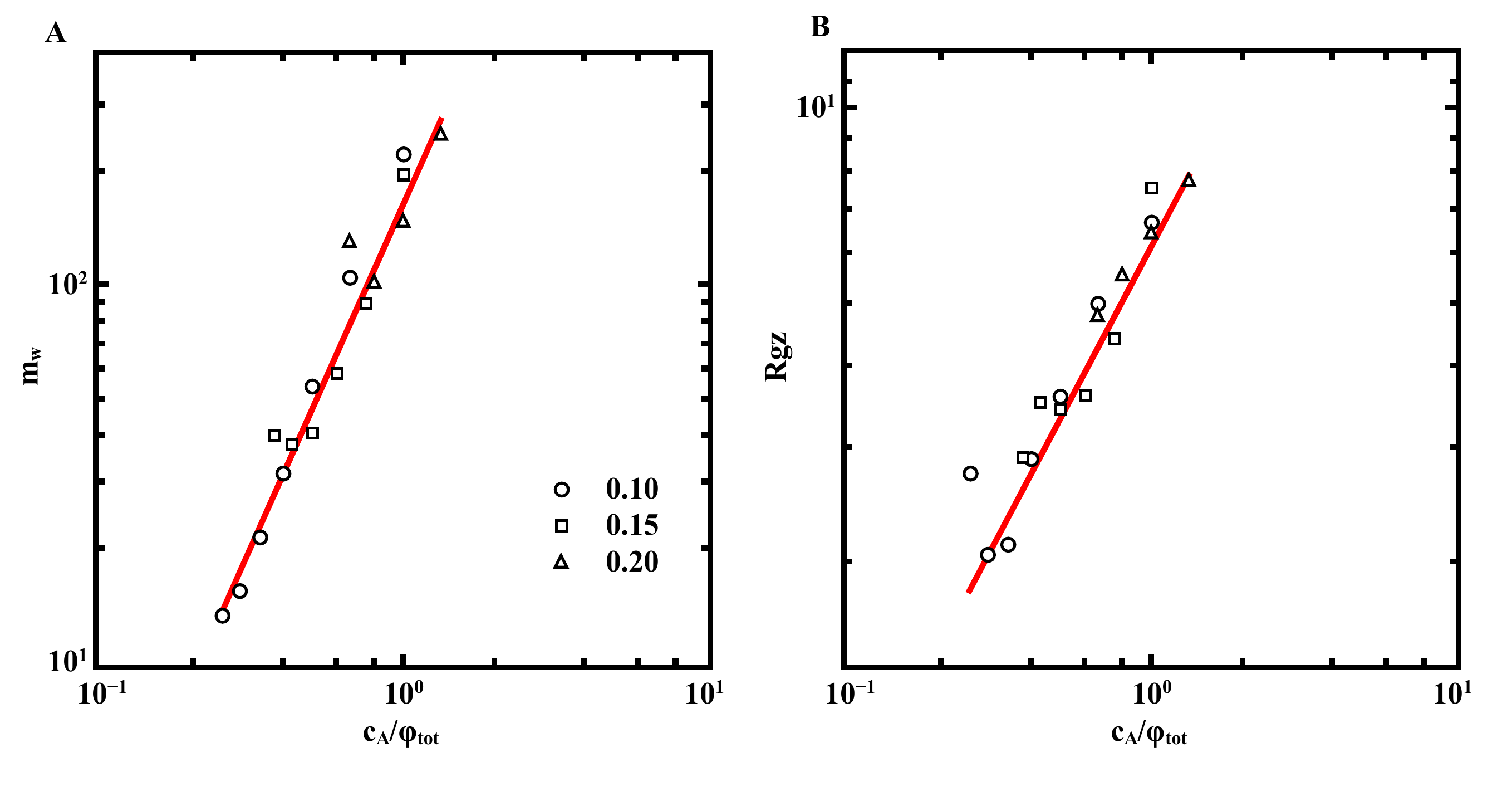}
 	\caption{{\bf(a)} The stationary value of $m_w$ for the $A$ particles is plotted as a function of $c_A/\phi_{tot}$ at different $\phi_{tot}$ for the values of $c_A$ as indicated in the figure. The solid line has a slope of $1.8$ consistent with the prediction of percolation theory. {\bf(b)} The stationary value of $z$ average radius of gyration is plotted against $c_A/\phi_{tot}$  for different $c_A$ as indicated in the figure at different $\phi_{tot}$ for the $A$ particles when it cannot percolate. The solid line has a slope of $0.88$ .}
 	\label{fig:7}
 \end{figure*}

	 In Fig.\ref{fig:7} we have plotted the stationary value of $m_w$ for the $A$ particles as a function of $c_A/\phi_{tot}$ for a range of $\phi_{tot}$ for three different $c_A$ as indicated in the figure. It seems that $c_A/\phi_{tot}$ is the parameter that determines the variation of stationary value of $m_w$ once $A$ particles are stuck inside the pores of $B$ particles. In the Fig.\ref{fig:3} we observe that as we go towards the percolation transition the stationary value of $m_w$ or the compressibility of the system keeps on increasing till we have an infinite spanning network or in other words $m_w$ diverges. We also observe that $m_w$ scales with $c_A/\phi_{tot}$ with a slope of $1.8$ see solid line in Fig.\ref{fig:7}a. It is a well established fact that $m_w \propto (p-p_c)^{\gamma}$ below the critical value $p_c$ according to percolation theory. This gives us a clear indication that if we consider  $(p-p_c) \propto c_A/\phi_{tot}$, may be $c_A/\phi_{tot}$ is the critical parameter for irreversible aggregation of binary system. We know that percolation theory predicts that the correlation length $\xi$ scale with $(p-p_c)^{\nu}$  were $\nu=0.88$ below the percolation threshold. It has already been shown that $Rg_{z} \propto \xi$, where $Rg_{z}=\sqrt{ \frac{\sum_{m=1}^{\infty} m^2 Rg^2}{ \sum_{m=1}^{\infty} m^2N(m)}}$ \cite{stauffer1979scaling}. In Fig.\ref{fig:7}b we have plotted $Rg_{z}$  with $c_A/\phi_{tot}$ and the solid line is given by the slope of $0.88$, which is in agreement with  the percolation theory. In Fig.\ref{fig:7} compressibility  and correlation length are showing universal behavior when plotted as function of $c_A/\phi_{tot}$ with the scaling law predicted by percolation theory, where by we can confirm that irreversible binary system comes under the universality class of percolation.
\section{Discussion}
	 Hecht et. al  have analyzed the fractal dimension of binary aggregating colloidal particle, where they have reported a $d_f=1.8$, which is predicted for $1$ component DLCA \cite{hecht2016kinetically}. With the experimental data of $m$ and $R_g$, they have reported it is very difficult to differentiate a slope of $1.8$ and $2$. Unfortunately they have not calculated $N(m)$ as a function of $m$, which we believe will give much more precise value of $d_f$, as shown in the present study. In the present work both the particles start aggregating simultaneously, while in  the experimental case  the aggregation of $A$ and $B$ particle starts at different times. It has already been shown that compared to a random distribution an aggregated system have more free space \cite{babu2008tracer}. Thus we believe that the diffusion and the kinetics of the system may be different, but there should not be much difference in the final  structure of these systems.
 
	 The appearance of bigel have also been reported by Varrato et. al, in reversible aggregation of binary colloidal system. They have reported $3$ distinct region, where both $A$ and $B$ percolated, only $B$ percolated and region where neither $A$ or $B$ percolated. In our case we only have $2$ regions  bigel or $1$ component gel, as expected from irreversible aggregation. For $\phi_{tot}>0.2$ our results are similar with the results of Varrato et. al, but for $\phi_{tot}<0.2$ they have reported $c_A$ much higher than what we have observed in our system. In the case of reversible aggregation it was shown that for low volume fraction only high attraction strength could mimic irreversible aggregation \cite{babu2006phase}. For smaller attraction the system undergoes spinodal decomposition for the $B$ particles, while the fraction of $A$ densifies which will require  a higher fraction in order to percolate.
 
	 In the present work we have shown that by controlling $\phi_{tot}$ and $c_A$ we are able to produce self similar clusters of particular size for the $A$ particles there by controlling the pore size of $B$ . This result can help in creation of new functional materials of a particular size simply by controlling the pores size created by the $B$ particles \cite{liu2006smart}. It will be interesting to further extend this study to include asymmetric particles, and study how the porosity of the corresponding gel varies. 
\begin{acknowledgments}

\end{acknowledgments}

\nocite{*}

\bibliography{references}
\end{document}